\documentclass[12pt,preprint]{aastex}
\usepackage{natbib}
\usepackage{graphicx, amsmath, amssymb}
\usepackage{graphicx}
\usepackage{subfigure}
\usepackage{epsfig}
\usepackage{color}
\usepackage{footmisc}

\begin{document}

\title{Disentangling AGN and Star Formation Activity at High Redshift Using \emph{Hubble Space Telescope} Grism Spectroscopy}

\author{Joanna S. Bridge\altaffilmark{1}, Gregory R. Zeimann\altaffilmark{1}, Jonathan R. Trump\altaffilmark{1,2}, Caryl Gronwall\altaffilmark{1}, Robin Ciardullo\altaffilmark{1}, Derek Fox\altaffilmark{1}, and Donald P. Schneider\altaffilmark{1} }
\affil{Department of Astronomy \& Astrophysics, The Pennsylvania State University, University
Park, PA 16802}
\email{jsbridge@psu.edu, grzeimann@psu.edu, jtrump@psu.edu, caryl@astro.psu.edu, rbc@astro.psu.edu, dfox@astro.psu.edu, dps7@psu.edu}
\altaffiltext{1}{Institute for Gravitation and the Cosmos, The Pennsylvania State University, University Park, PA 1602}
\altaffiltext{2}{Hubble Fellow}

\begin{abstract}
Differentiating between active galactic nuclei (AGN) activity and star formation in $z \sim 2$ galaxies is difficult because traditional methods, such as line ratio diagnostics, change with redshift while multi-wavelength methods (X-ray, radio, IR) are sensitive to only the brightest AGN\null. We have developed a new method for spatially resolving emission lines in \emph{HST}/WFC3 G141 grism spectra and quantifying AGN activity through the spatial gradient of the [O~III]/H$\beta$ line ratio. Through detailed simulations, we show that our novel line-ratio gradient approach identifies $\sim 40\%$ more low-mass and obscured AGN than obtained by classical methods. Based on our simulations, we developed a relationship that maps stellar mass, star formation rate, and measured [O~III]/H$\beta$ gradient to AGN Eddington ratio.  We apply our technique to previously studied stacked samples of galaxies at $z\sim2$ and find that our results are consistent with these studies. Using this gradient method will also be able to inform other galaxy evolution science, such as inside-out quenching and metallicity gradients, and will be widely applicable to future spatially resolved \emph{JWST} data.
\end{abstract}

\section{Introduction}
The study of high-redshift galaxies has been hampered by the fact that their strongest emission lines, such as $H\alpha$, H$\beta$, [O~II] $\lambda 3727$, and [O~III] $\lambda 5007$, are shifted out of the optical bands and into the near-infrared, where atmospheric opacity, high background, and telluric emission compromise ground-based observations. The infrared grisms of the \emph{Hubble Space Telescope} (\emph{HST}) give unique access to this wavelength regime.  Because the \emph{HST} grisms are slitless, they offer unprecedented multiplexing ($\gtrsim 50$ galaxies per arcmin$^2$) for the study of high-redshift emission-line science. In particular, the near-IR 3D-HST (GO-12177, 12328; PI: P. van Dokkum) and AGHAST (GO-11600; PI: B. Weiner) grism surveys, with the G141 grism on Wide Field Camera 3 \citep[WFC3;][]{kimble2008} on \emph{HST}, has provided a sample of over 7000 galaxies between $1 < z < 3.5$ \citep{brammer2012}.

Unfortunately, our ability to interpret the rest-frame optical emission lines is compromised by the fact that our knowledge of emission lines in the nearby universe does not always translate to high-redshift. At $z > 1$, typical galaxies have more star formation \citep{madau2014}, harder ionizing radiation \citep{kewley2015}, and smaller physical sizes \citep{vanderwel2014} than their local counterparts.  As a result, diagnostic diagrams, such as that developed by \citet[BPT;][]{baldwin1981}, can change with redshift \citep[e.g.,][]{kewley2013b, steidel2014}, complicating the study of high-redshift star-forming systems. For example, distinguishing an active galactic nuclei (AGN) from normal star-forming galaxies can become problematic, especially when dealing with a low-mass or obscured object. As the Eddington ratio ($\lambda_{\rm Edd} = L_{bol}/ L_{\rm Edd}$) declines, radio and IR methods become less efficient at AGN selection because they are increasingly confused with emission from star formation at the high star formation rates typical of $z>1$ galaxies \citep{donley2012, kimball2011}. X-ray selection, typically the most reliable way to identify unobscured AGNs, is biased against AGNs in low-mass galaxies \citep{xue2010, aird2012}, and a large fraction ($\gtrsim$ 50\%) of AGN activity may be obscured in X-rays \citep[e.g.,][]{buchner2015}.  Emission-line ratios offer promise for finding X-ray-obscured AGN, but even this signature can be diluted by emission associated with star formation, especially in low-mass galaxies \citep{moran2002, trump2015}.

At low redshifts, AGN have harder ionizing radiation than typical H II regions, resulting in higher ratios of [O~III]/H$\beta$ and [N~II]/H$\alpha$  in the classic BPT diagram.  At $z \sim 2$, however, studies \citep[e.g.,][]{kewley2013b, juneau2014} are finding that star-forming galaxies can have a large range in excitation (as defined by [O~III]/H$\beta$ and [N~II]/H $\alpha$) that may mimic AGN\null. There are a number of possible reasons for this result including harder ionization fields \citep[e.g.,][]{steidel2014}, denser interstellar media and higher ionization parameters \citep[e.g.,][]{brinchmann2008, kewley2013a, shirazi2014}, and/or shifts in the abundance ratio of nitrogen to oxygen \citep[e.g.,][]{masters2014}. In addition, investigation of small samples of spatially resolved galaxies have posited that there is contamination by obscured or low-mass AGN in regular star-forming galaxies, resulting in a shift in the observed line ratios \citep{wright2010, trump2011}. Given these variations, it would be useful to understand how 2D grism spectroscopy can be used to probe the parameter space of properties that affect emission line diagnostics.

Spatially resolved spectroscopy has shown great utility in understanding the H$\alpha$ star formation rate profiles \citep{nelson2012, nelson2013, nelson2015, wuyts2013}, kinematic profiles \citep{forster2009, wisnioski2015}, and dust extinction profiles \citep{nelson2016} of high-redshift galaxies.  But large studies of emission-line ratios in high-redshift galaxies have been limited to integrated line ratios \citep[e.g., MOSDEF;][]{kriek2015, coil2015}. The HST/WFC3 grism, however, increases the number of spatially resolved high-redshift galaxies by an order of magnitude.  Our investigation seeks to complement this ground-based effort with a large sample of spatially resolved emission lines that offers a unique handle on AGN identification using the different spatial extent of AGN and galaxies.

In this work, we use simulations to investigate the use of spatially distinguished emission-line ratios for AGN selection, quantifying the technique's sensitivity as a function of AGN accretion rate and galaxy properties. In \S 2, we present our simulated galaxies and the simulated grism observations.  We discuss the limitations of traditional AGN identification methods in \S 3 and in \S 4, we discuss how we improve upon these methods by collapsing the 2D grism spectra to derive emission line information about each simulated galaxy.  We apply this method to data in \S 5 and we discuss the implications of these results for grism surveys of the high-redshift universe in \S 6.  We conclude with \S 7 and examine how these data can be used to refine future galaxy studies.

In this work, we adopt the standard $\Lambda$CDM cosmology, with $H_0=70$ km s$^{-1}$ Mpc$^{-1}$, $\Omega_M=0.3$, and $\Omega_\Lambda=0.7$ \citep{komatsu2011}.

\section{Galaxy and AGN Simulations}
Our goal is to measure the utility of spatially determined emission-line ratios of $z>1$ galaxies for AGN identification and characterization. The key issue is one of contrast: The relative brightness of AGN light with respect to the extended galaxy emission associated with star formation or shocks varies by many orders of magnitude. To address this issue, we created detailed simulations of \textit{HST}/WFC3 grism observations by modeling a point-source AGN with an extended star-forming host galaxy. The simulations include a range of AGN accretion rates and galaxy properties. Building on the classic integrated line-ratio diagnostics, we measure the [O~III]/H$\beta$ \textit{gradient} in the galaxy from the combination of the central AGN and extended galaxy emission. The [O~III]/H$\beta$ gradient is defined as
\begin{equation}
 \textrm{log([O III]/H}\beta)_{\textrm{inner}} - \textrm{log([O III]/H}\beta)_{\textrm{outer}}
 \end{equation}
We examine the values of this [O~III]/H$\beta$ gradient over a wide range of simulated galaxy and AGN parameters, seeking to understand how the measured gradient corresponds to galaxy and AGN properties.
 
\subsection{Simulations}

We focus our attention on mimicking real observations of the well-observed CANDELS fields \citep[Cosmic Assembly Deep Near-IR Extragalactic Legacy Survey;][]{grogin2011, koekemoer2011}.  The 3DHST \citep{brammer2012} and AGHAST \citep{weiner2014} programs covered the five CANDELS fields using two orbits of G141 grism.  However, the rather short total exposure time ($\sim90$ minutes) is not sufficient for the study of moderate-luminosity galaxies; this is particularly true for the low-mass, low-Eddington ratio parameter space we intend to explore. We therefore stacked twenty systems grouped by physical properties for an effective exposure time of 40 orbits, or roughly 30 hours. Stacked galaxy spectra also allow us to assume average galaxy properties, avoiding the broad diversity of individual galaxy morphologies.  Almost all previous observational efforts to spatially examine WFC3 grism spectra have used stacked spectra \citep[e.g.,][the only non-stacked exceptions use highly lensed galaxies]{trump2011, trump2014, nelson2012, nelson2013, nelson2015, nelson2016, whitaker2013, wuyts2013}.

For the simulated observations, we use the G141 slitless grism, with a spectral range of $1.08 < \lambda < 1.68$ $\mu$m and resolution of $R = 130$.  The inherent spatial pixel scale of the grism is $0\farcs12$, which we then drizzled to a finer scale of $0\farcs064$.  For the main set of simulations, the redshift is set at a constant $z = 1.8$, which places the [O~III] and H$\beta$ emission lines approximately at the center of the G141 grism. We also simulate $z = 1.4$ and $z = 2.2$ to demonstrate that our simulations remain appropriate for interpreting observations at the edges of the G141 wavelength range. Therefore, our results are broadly applicable to the broader $1.3<z<2.4$ redshift range over which the H$\beta$ and [O~III] lines are observed in the G141 grism. Finally, we explore the effects of different galaxy morphologies by simulating changes is ellipticity and light profile in galaxies. While ellipticity and diverse morphologies are likely to be averaged out in the stacking of real galaxy images, testing the effects of other profiles is important to ensure the robustness of the method.

The spatially resolved line ratios in our simulated galaxies are produced by a combination of extended star formation and nuclear AGN:  While the H$\beta$ equivalent width and [O~III]/H$\beta$ ratio produced by star formation is a function of the galaxy's stellar mass and specific star formation rate (sSFR), the [O~III] equivalent width of the nuclear AGN is determined by its Eddington ratio (i.e., specific accretion rate).  Thus, to model our $z = 1.8$ galaxies, we need to consider two components:  a (face-on) galactic disk, which we model with an exponential profile using IRAFÕs {\tt mkobject} algorithm, and a point-source AGN, which is simulated via the instrumental point spread function (PSF)\null.   An illustration of these components is shown in the first column of Figure~\ref{grism}. Our simulations then consist of a grid of direct and dispersed images with stellar mass, sSFR, and AGN Eddington ratio as the input model parameters.

Using the \texttt{EzGal} Python program \citep{mancone2012}, we convert the input stellar mass to the galaxy spectral continuum in the observed G141 wavelength range.  In this process, we employed the stellar population synthesis models of \citet{bruzual2003} using solar metallicity and a \citet{chabrier2003} initial mass function (IMF). By projecting a spectral energy distribution (SED) through a filter response curve at a given redshift, \texttt{EzGal} calculates the absolute AB magnitude of a galaxy for a particular stellar population synthesis (SPS) model \citep{mancone2012}. 

We also took into account physical sizes of the galaxies. Variation arises from the spread in the size-mass relation given by \citet{vanderwel2014},
\begin{equation}
R_{\textrm{eff}} = \textrm{A}\cdot m_*^{\alpha}
\end{equation}
where for $z=1.8$, log A = $0.6\pm0.01$, $\alpha=0.23\pm0.01$, and the width of the distribution is $\sigma$log(R$_{\textrm{eff}}$) = $0.1\pm0.01$ (Equation 3, Table 1; \citealt{vanderwel2014}).  Here, $R_{\textrm{eff}}$ is the effective, or half-light, radius of the galaxy, in kpc, and $m_*$ is the stellar mass. We assign each galaxy a size using Equation 1, drawing from a normal distribution with a 0.18 dex standard deviation.

The galactic [O III]/H$\beta$ ratio is complicated as it is a function of nebular density, ionization parameter, ionizing spectrum, and metallicity ($Z$).  We use the input stellar mass and sSFR to calculate  an [O III]/H$\beta$ ratio via the mass-SFR-$Z$ relation of \cite{mannucci2010} and the strong line ratio relations of \cite{maiolino2008}. The values cover the range 0 $<$ log([O~III]/H$\beta$) $<$ 0.7 over the mass and sSFR parameter space of the simulations. We then converted our star formation rates to Balmer line flux using the relation of \citet{kennicutt1998} and combined these measurements with
{\tt EzGalÕs} continuum magnitude to estimate equivalent width.   We note that both the
\citet{mannucci2010} and \citet{maiolino2008} relations are derived from SDSS
galaxies, and studies have shown that as the redshift increases, line ratio values may also
increase \citep[e.g.,][]{shapley2005, erb2006, kriek2007, kewley2013a}. Our simulations do not include this effect.

To characterize the AGN, we need to convert from Eddington ratio to an [O~III] equivalent width from the AGN\null.  By assuming $M_{\textrm{BH}}/M_* \sim 0.001$ (Haring \& Rix et al.\ 2004), we obtain a black hole mass ($M_{\rm BH}$) given the stellar mass of the galaxy.  Combining $M_{\rm BH}$ with the Eddington ratio, we have the bolometric luminosity of the AGN\null. We then modeled the AGN emission with [O~III]/H$\beta = 1.0$, noting that the ratio does not change with metallicity as it is virtually saturated \citep{kewley2013a}. This AGN ratio value is at the high end of typical AGN [O~III]/H$\beta$ ratios and in reality may be lower and vary as a function of ionization parameter \citep{richardson2014}. To determine the AGN [O~III] luminosity, we adopt the method prescribed by \citet{trump2015}, which uses a power-law fit to the \citet{lamastra2009} luminosity-dependent bolometric corrections of
\begin{equation}
\frac{L_{\textrm{bol}}}{10^{40}\hspace{2pt}{\textrm{erg s}^{-1}}} = 112 \left( \frac{L[\textrm{OIII}]}{10^{40}\hspace{2pt}{\textrm{erg s}^{-1}}} \right)^{1.2}
\end{equation}
(Equation 9; \citealt{trump2015}).  In this way, we convert from a given combination of bolometric luminosity and stellar mass to an AGN [O~III] luminosity.

Once we fully described each galaxy and AGN combination, each grid point of galaxy properties was ``observed" 100 times using the \texttt{aXeSim}\footnote[4]{http://axe.stsci.edu/axesim/} software package. These 100 individual simulations were placed evenly over the entire detector, thereby averaging over any field dependencies that might arise due to object placement in the detector. The results of an observation are shown in the second column of Figure~\ref{grism}. Each of the 100 simulated galaxies had a radius drawn from the size distribution for the given stellar mass being simulated, so as to encompass possible variations in the observed 2D spectrum. Using the grid points indicated in Table 1, our final galaxy dataset consisted of 46,200 simulations.

\subsection{Discussion of Simulations}
In these simulations, we have not included the effects of extinction. Extinction will decrease the line equivalent widths, since lines and stellar continuum have different extinction \citep{calzetti2000}.  However, it is unlikely to affect the \textit{contrast} between the AGN and star formation, since extinction is likely to affect both AGN and star formation equally.  The AGN narrow line region is likely $\sim1$ kpc in size \citep[e.g.,][]{bennert2006}, and, therefore, is likely to have the same extinction as the star-forming region.  \citet[][see their appendix]{trump2015} also demonstrate that $z\sim0$ narrow-line AGN have the same dust distribution as star-forming galaxies.

It is possible that the extended line ratios associated with star formation are not constant with radius, due to gradients in metallicity or ionization.  Metallicity gradients of both local and distant galaxies are typically negative, with higher metallicity (and consequently lower [O~III]/H$\beta$) in the center \citep[e.g.,][]{swinbank2012, jones2013, sanchez2014}; this would actually increase the contrast between a nuclear AGN and star formation. Shocks can potentially cause regions of higher ionization but they are unlikely to produce [O~III]/H$\beta$ ratios like an AGN\null.  An inclusion of nuclear shocks would lower the AGN threshold of [O~III]/H$\beta$, although not significantly \citep{allen2008}. Therefore, in our current study, we neglect the effects of abundance gradients for both simplicity and due to a lack of high-redshift observational constraints over a wide range of stellar masses and star-formation rates. Current observations of [O~III]/H$\beta$ are still limited, but \cite{whitaker2014} found [O~III]/H$\beta$ ratios to be relatively flat across 6 kpc at $z = 1.7$ for a single lensed galaxy, in support of our simplifying assumption.

Another caveat is the use of \cite{mannucci2010} and \cite{maiolino2008}, in combination with SFR and stellar mass, to predict the underlying [O~III]/H$\beta$ ratios of our model galaxies.  If these relations do not hold at higher redshift, as some studies suggest \citep[e.g.,][]{sanders2015, gebhardt2015}, then this would affect our measured [O~III]/H$\beta$ gradients.  This systematic effect is difficult to quantify as the evolution of [O~III]/H$\beta$ as a function of redshift, stellar mass, and sSFR has not yet been determined, but it crucially affects the assumed star-forming line ratios.

Finally, it should be noted that our simulations are best suited to stacked galaxy ensembles.  At the level of two \emph{HST} orbits, performing this analysis on single (unlensed) galaxies is generally not possible. Since stacked data average over many morphological parameters, we are able to simplify this part of the simulations. For example, the asymmetry of a galaxy image in the dispersion direction would complicate the annular collapse method, since we collapse only the blue side of H$\beta$ and red side of [OIII].  Another example is that of galaxy concentration: A higher-S\`ersic profile would decrease the contrast between nuclear AGN and extended star formation. Using a disk profile is a legitimate assumption for considering the AGN content of typical star-forming galaxies, but galaxies above or below the star-forming ``main sequence" are more likely to have more compact morphologies \citep{wuyts2011}. The diversity of individual galaxy shapes and sizes that would be an issue for single-galaxy analysis is mitigated by our focus on stacked data.

\section{Limitations of Integrated Line Ratios}
The classical method used to differentiate AGN from star formation in galaxies is to employ integrated emission line ratios, either compared to each other \citep[e.g.,][]{baldwin1981} or to a given physical parameter, such as stellar mass \citep[e.g.,][]{juneau2011}.  This method depends on the AGN dominating the line fluxes of the entire galaxy; while this may be true of massive galaxies with luminous AGN, there is potential to overlook low-mass or obscured AGN\null. Our inability to probe the low-mass portion of AGN parameter space has ramifications for our understanding of the black hole occupation fraction and black hole seeds \citep[e.g.,][]{moran2014, trump2015}.

To illustrate the limitations of integrated line ratios, we extract an integrated 1D spectrum from the full 2D grism observation. We then fit three Gaussians (H$\beta$, [O~III]$\lambda\lambda4959, 5007$) plus a polynomial continuum to each galaxy's spectrum to measure its total observed line flux.  For the 100 simulations at each grid point, we calculate the ``completeness"; that is, the percentage of simulations in which we identify the AGN\null.  This threshold for integrated line ratios is a function of the stellar mass of the galaxy, taken from the Mass-Excitation (MEx) diagram of \citet{juneau2011}.

This completeness is shown in Figure~\ref{hex_diff_tot}. Each hexagon is shaded by the completeness for each grid point.  As expected, using integrated line ratios works extremely well when the Eddington ratio of the galaxy is large and the sSFR is low, thereby allowing the AGN emission to dominate the galaxy's emission.  This method begins to fail, however, for low-luminosity AGN, independent of the sSFR of the galaxy.  In addition, the use of integrated line luminosities for low-Eddington ratio, high-sSFR galaxies is problematic because the emission from star formation begins to overtake the AGN line emission.

Another way to view the MEx's inability to identify all AGN is presented in Figure~\ref{Mex}.  The integrated line ratios from the simulations are plotted as a function of stellar mass with green circles, while the black lines represent the empirical separation between AGN and star formation \citep{juneau2014}. Every bin from medium to low stellar mass has AGN that fall below this dividing line; this is especially true at the lowest masses where the cut-off in [O~III]/H$\beta$ is high. Our goal is to augment this parameter space by developing a method to recover more of the AGN that failed to be identified using the MEx diagram.

\section{AGN Selection from Spatially Resolved Line Ratios}
Integrated line luminosities are essentially a blunt tool for studying emission from galaxies.  What is needed is a more nuanced method for decomposing the emission into its individual sources. The 2D grism spectra produced by \emph{HST} provide such a tool as it allows compact nuclear emission to be at least partially resolved from its surroundings.  To do this, we divide a grism spectrum into two regions -- the nuclear region that is likely to contain emission from a black hole, and an extended disk, whose light is dominated by emission from star formation.This technique was first presented by \citet{trump2011}.  Here, we use simulations to test the validity of this method as well as its strengths when compared to traditional AGN identification methods.

\subsection{Measuring the Emission Lines}
\begin{figure}
\centering
\scalebox{0.5}
{\includegraphics{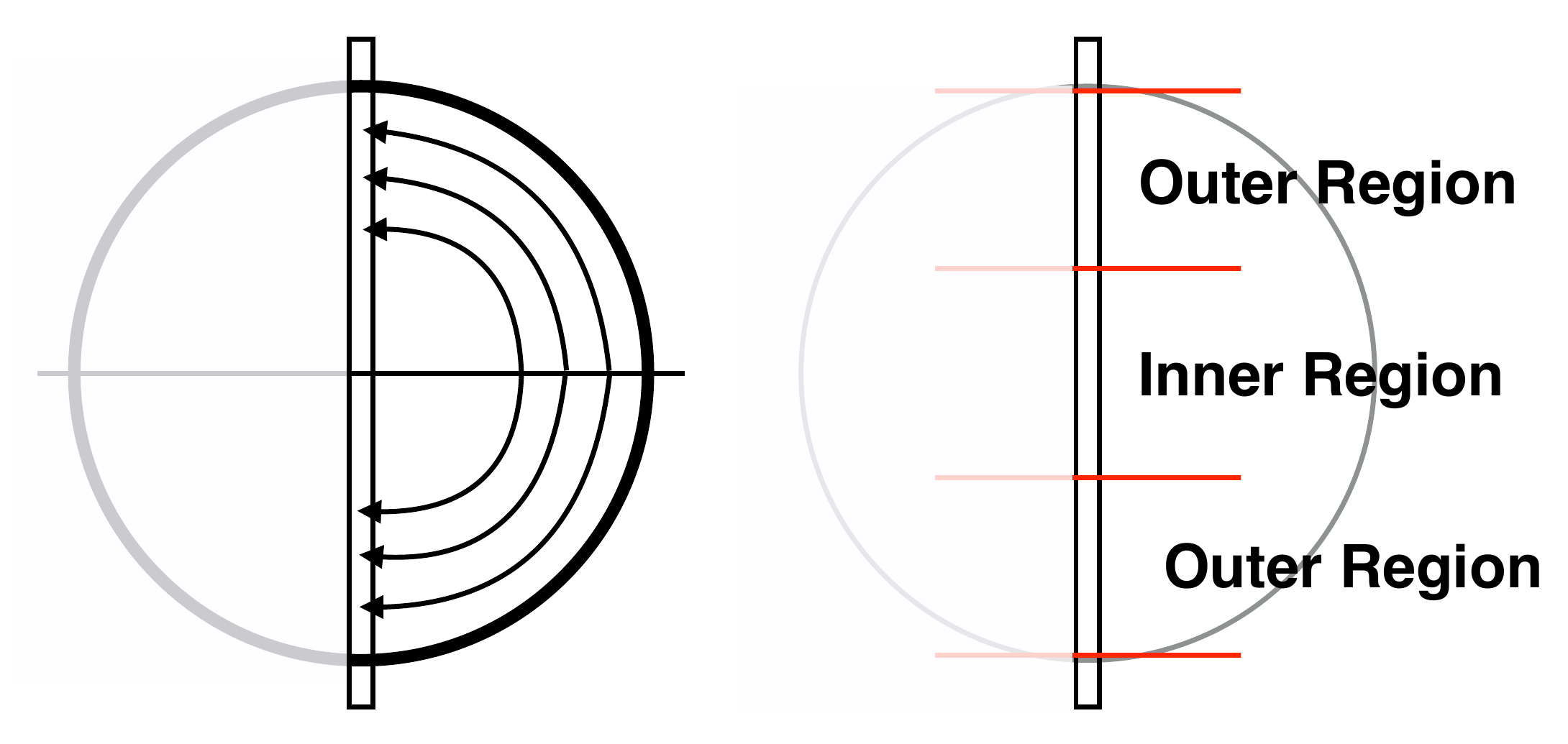}}
\centering
\caption{An illustration of how the collapsing technique works for the right-hand side of the [O~III]$\lambda$5007 line. The flux is summed along lines of constant radii, so that each pixel in the final column of pixels is the sum of the flux the annulus collapsed. The resulting single column can then be separated into the inner and outer regions of the galaxy. We repeat this process for the left-hand side of the H$\beta$ line, from which we calculate an accurate [O~III]/H$\beta$ ratio for each region.}
\label{schematic}
\end{figure}

\begin{figure}
\centering
\scalebox{0.5}
{\includegraphics{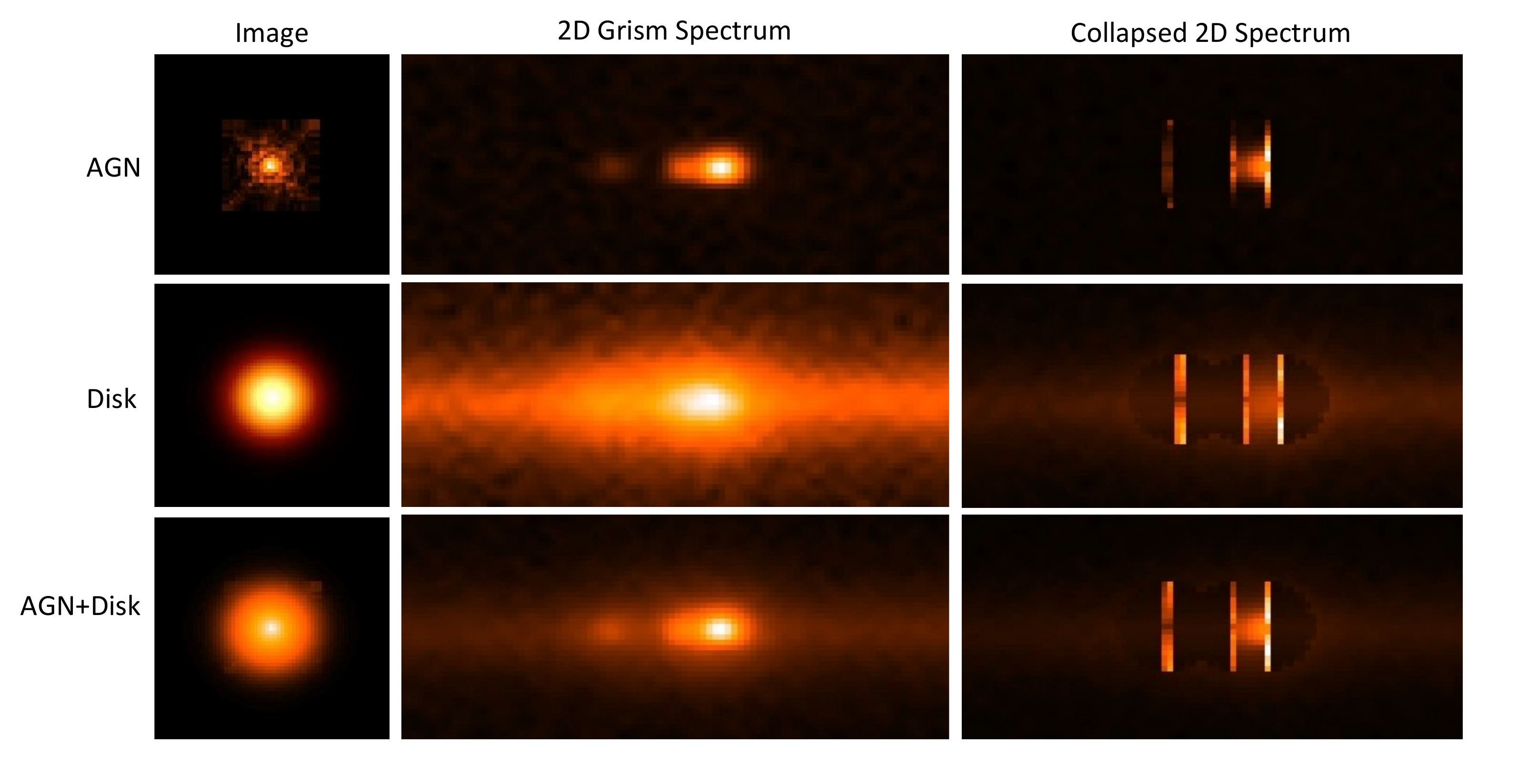}}
\centering
\caption{\emph{Top row}: An AGN modeled as a PSF (left), a simulated grism observation of the
AGN with an [O~III] and H$\beta$ emission (center), and the collapsed continuum-subtracted 2D spectrum (right). By adding (fractional) pixels along annuli of constant radii, we have confined the spatial extent of the galaxy to the cross-dispersion direction. \emph{Second row}: Same as the first row, but with a star formation modeled as an exponential disk, and an input spectrum of [O~III] and H$\beta$ typical of a star-forming galaxy. \emph{Third row}: Same as the first row but with the AGN and the exponential disk combined.}
\label{grism}
\end{figure}

\begin{figure}
\centering
\begin{subfigure}
	\centering
	\scalebox{0.39}
	{\includegraphics{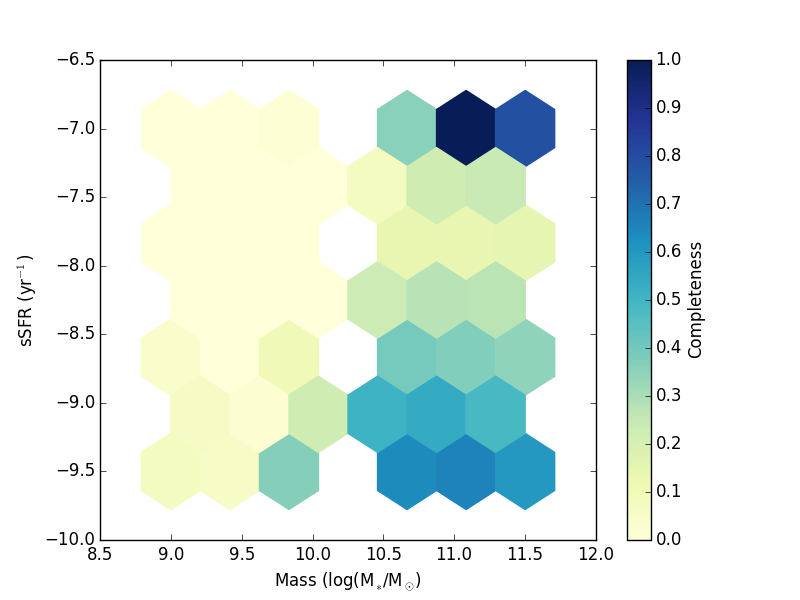}} \\
\end{subfigure}
\begin{subfigure}
	\centering
	\scalebox{0.39}
	{\includegraphics{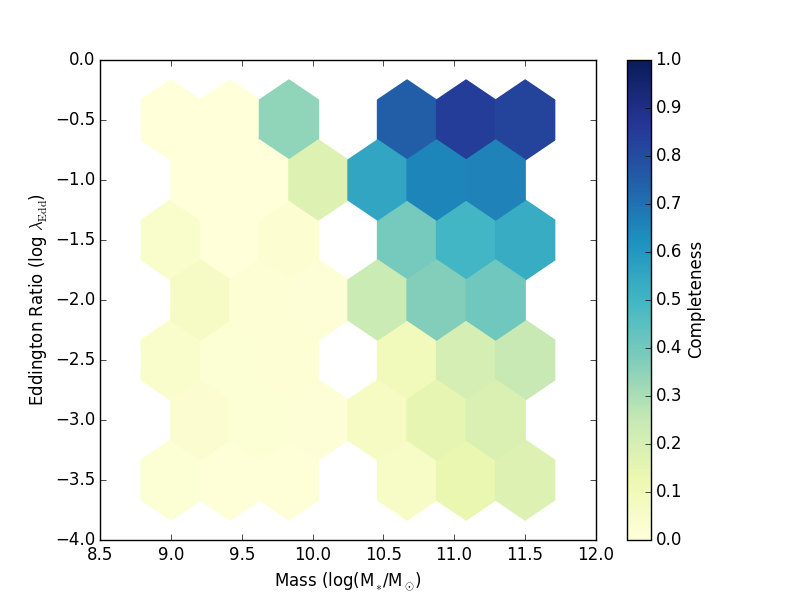}}
\end{subfigure}
\begin{subfigure}
	\centering
	\scalebox{0.39}
	{\includegraphics{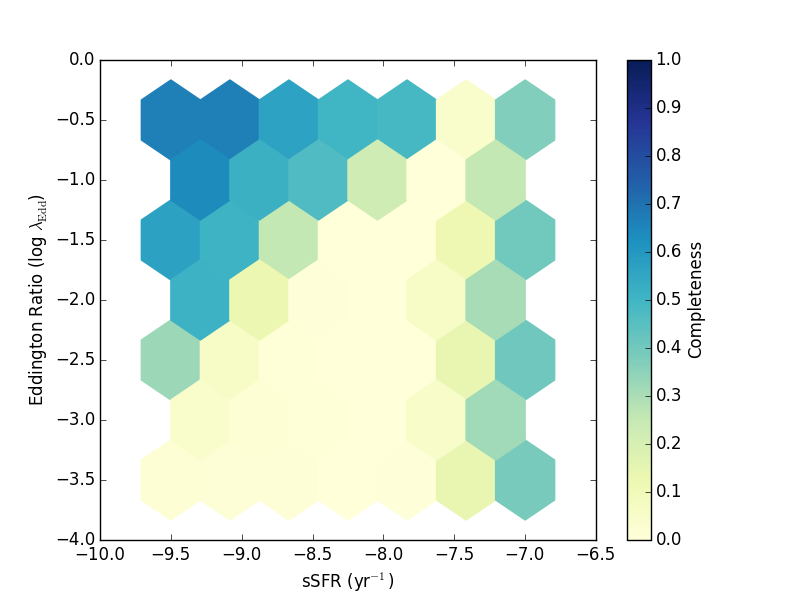}}
\end{subfigure}
\caption{Simulated parameters for the total integrated flux line ratios, shaded by completeness using the emipirical cut of \citet{juneau2014} (see Figure~\ref{Mex}). Integrated line ratios are successful at finding luminous AGN in massive, weakly star-forming galaxies but are unable to identify AGN in low-mass and rapidly star-forming hosts.}
\label{hex_diff_tot}
\end{figure}

\begin{figure}
\centering
\scalebox{0.8}
{\includegraphics{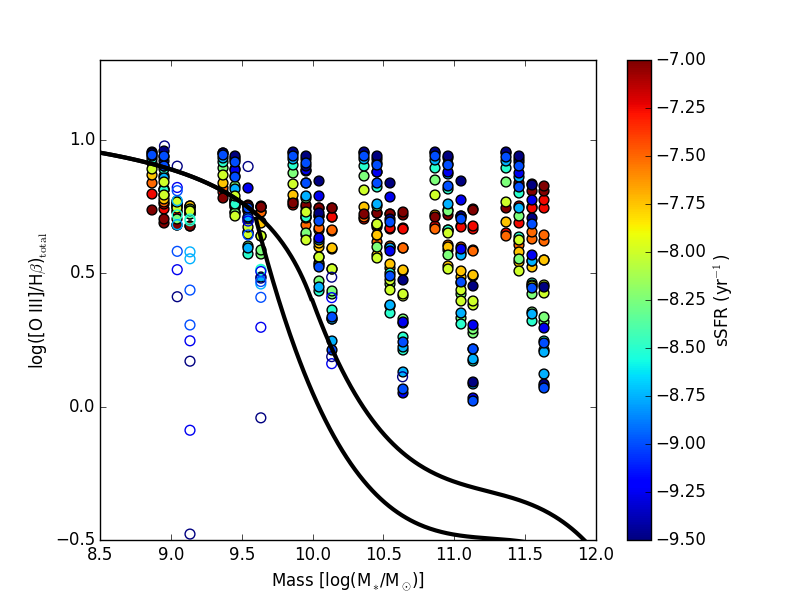}}
\centering
\caption{Mass-Excitation diagram for the simulated integrated log([O~III]/H$\beta$) ratios shaded by sSFR. Four groups of Eddington ratios are offset in mass, from low to high. The galaxies with low signal-to-noise detections are plotted with empty circles. The black line is the empirical cut between AGN and star formation \cite{juneau2014}, accounting for the luminosity limit and redshift evolution of $L_{[\rm{O~III]}}$. At lower stellar masses, it becomes more difficult to identify AGN as star formation dilutes the AGN signal.}
\label{Mex}
\end{figure}

The slitless grism data are spatially extended in both the dispersion and cross-dispersion dimensions, so it is nontrivial to extract nuclear and extended 1D spectra. The low spectral resolution (R = 130) and lack of a slit in the grism means that extended features in the dispersion direction are due to a galaxy's physical size rather than its velocity extent. Rather than deconvolving in two dimensions (dispersion and cross-dispersion), we simplify the spectra by ``collapsing" each emission line along constant radii, reducing the spatial extent into the cross-dispersion direction only.  In other words, the flux is collapsed annularly, adding fractional pixels along constant radii to a single column in the cross-dispersion direction, weighting each pixel by the inverse variance. Errors in the collapsed pixels are computed analytically with the total error given as the inverse sum of the weights: $\sigma^2 = 1 / \Sigma(f_i/\sigma_i^2)$, where $f_i$ is the pixel fraction and $1/\sigma^2$ is the weight. An illustration of this process is shown in Figure~\ref{schematic}, and the results are illustrated in the third column of Figure~\ref{grism}.  

In the simulation, the emission-line centers are well-known by definition.  This will be true for most applications of observational data, as it is difficult to measure a line ratio gradient without accurately knowing the line centers.  Our definition of the nuclear and extended apertures means that line-center errors of $<$1 pixel (i.e., typical grism-based redshifts with visible emission lines) will produce no changes in the observed line-ratio gradient.  Larger line-center errors will effectively dilute and reduce the observed line-ratio gradient, so long as these errors are symmetric among the galaxies being stacked.

After identifying the line centers, we measure and subtract the continuum by fitting to it a polynomial. Then, because the [O~III] doublet is blended in the grism spectra, and because H$\beta$ overlaps with [O~III] $\lambda 4959$, we assume line symmetry and collapse only one half of each emission line. We therefore use the right-hand side of the [O~III]$\lambda$5007 line and the left-hand side of the H$\beta$ line to calculate the [O~III]/H$\beta$ ratio for each  galaxy. We then divide that column of pixels into a nuclear region (inner three pixels, $R \lesssim 0.6$ kpc) and an extended region (outer apertures of pixels 3-5, $1.3<R<2.3$ kpc).  Finally, we calculate the inner and outer [O~III]/H$\beta$ ratio for each simulation. We implement a signal-to-noise cut of S/N $> \sqrt{3}/2$, noting that \citet{juneau2014} demonstrated that imposing this cut on the line ratios (rather than on individual lines) produces a reliable census of emission-line galaxies. This slightly reduces the number of useable galaxies from the original 100 simulated galaxies in bins with weak emission lines (typically low mass, low sSFR, and low Eddington ratio). This entire process allows us to calculate a line ratio \emph{gradient} for each simulated galaxy.  Therefore, instead of being limited to the total line luminosity, we have the ability to quantify the difference in emission from nuclear and extended regions of the galaxy. 

 \subsection{Determining the AGN Identification Threshold}
Once we have a line ratio gradient for each galaxy stack, the next step is to determine the threshold for AGN identification.  This decision requires us to know the completeness in terms of measured gradients when AGN are present, as well as the reliability in terms of the measured gradients when no AGN are present in the simulations.  To address the latter concern, we re-simulated our galaxies, but this time with only a star-forming disk and no AGN, removing the Eddington ratio dimension of the original parameter space.

\begin{figure}
\centering
\scalebox{0.8}
{\includegraphics{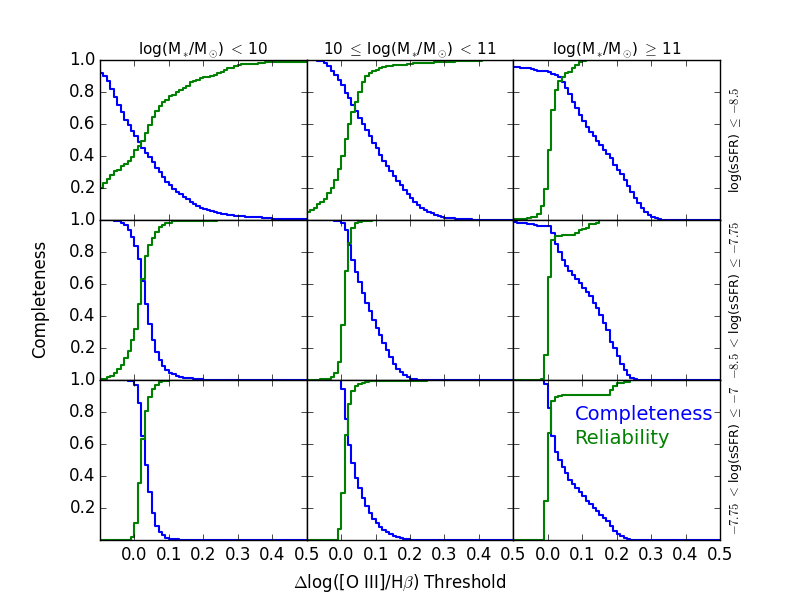}}
\centering
\caption{Cumulative distributions of AGN completeness (blue) and reliability (green) for our simulated galaxies. The reliability is determined by how often no AGN was found in the simulations with no AGN present. To optimize AGN identification, the false positives must be minimized while maximizing the completeness. We use the conservative threshold value of $\Delta$log([O~III]/H$\beta$) = 0.1 for all galaxies in our sample.}
\label{thresh}
\end{figure}

The best way to determine an appropriate gradient cutoff value is to maximize the completeness of AGN identification within the sample while minimizing the number of false positive identifications in the simulations with no AGN\null. Figure~\ref{thresh} displays the cumulative distributions of completeness and reliability, divided into three mass and three sSFR bins. As can be seen in the various panels, the best threshold gradient value can shift depending on the combination of stellar mass and sSFR.  While moving threshold may be useful for certain purposes, we choose the conservative value of $\Delta$log([O~III]/H$\beta$) = 0.1 for all galaxies in our sample.

\begin{figure}
\centering
\scalebox{0.8}
{\includegraphics{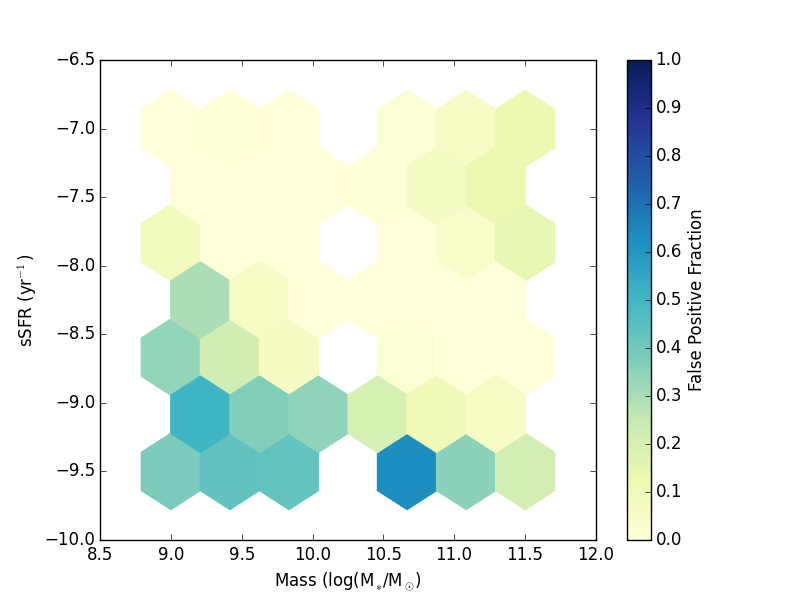}}
\centering
\caption{Specific SFR vs.\ stellar mass for the simulated galaxies with no AGN present, shaded by the rate of false AGN detections using the AGN identification threshold of $\Delta$log([O~III]/H$\beta$) = 0.1. The false positive rate is extremely low, although it becomes less reliable at lower masses where the signal-to-noise is lower.}
\label{false_pos}
\end{figure}

Using $\Delta$log([O~III]/H$\beta$) = 0.1 for our simulations, the false positive rates for our simulations are shown in Figure~\ref{false_pos}. Here we have displayed the number of AGN per grid point of the parameter space of the simulations with no AGN present.  The false positive rate is quite low across the board. This figure emphasizes the validity of using line ratio gradients to differentiate between AGN and star formation, showing that this technique is not prone to misidentification given our assumptions.

\subsection{Line Ratio Gradient Results}
\begin{figure}
\centering
\scalebox{0.8}
{\includegraphics{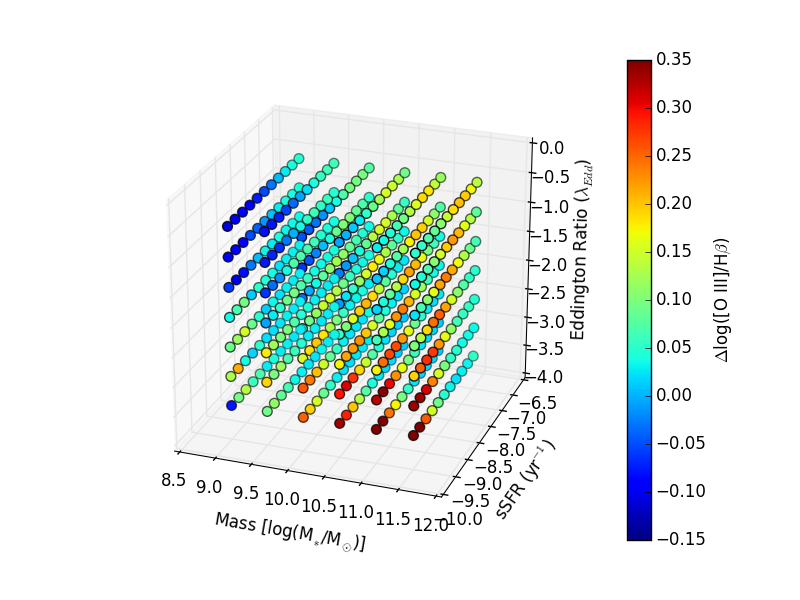}}
\centering
\caption{Gradient of log([O~III]/H$\beta$) as a function of stellar mass, sSFR, and Eddington ratio. The measured line-ratio gradient is highest in massive, low-sSFR galaxies with rapidly accreting AGNs, but it is an effective means of detecting an AGN over most of this phase space.}
\label{scatter3D}
\end{figure}

Figure~\ref{scatter3D} displays the $\Delta$log([O~III]/H$\beta$) gradients for our simulations as a function of input sSFR, Eddington ratio, and stellar mass grid points in our simulation matrix. The points are shaded by the median $\Delta$log([O~III]/H$\beta$) line ratio gradient for the simulations at each locus of the AGN parameter space.  The larger the gradient, the more apparent the AGN\null. From the figure, it is clear that using line ratio gradients to differentiate between disk star formation and nuclear AGN emission is effective in most combinations of sSFR and Eddington ratio. Discrimination becomes more difficult at lower mass but remains more successful than traditional methods.  This is true even when the sSFR is high and the Eddington ratio is low.  In this scenario, the emission from the star formation overwhelms the nuclear emission, causing the gradient to decrease.  

\begin{figure}
\centering
\begin{subfigure}
	\centering
	\scalebox{0.39}
	{\includegraphics{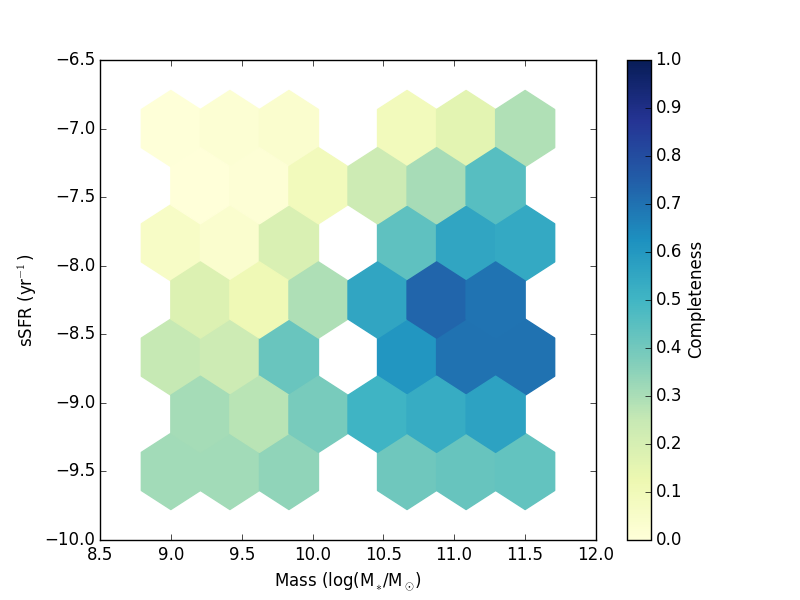}} \\
\end{subfigure}
\begin{subfigure}
	\centering
	\scalebox{0.39}
	{\includegraphics{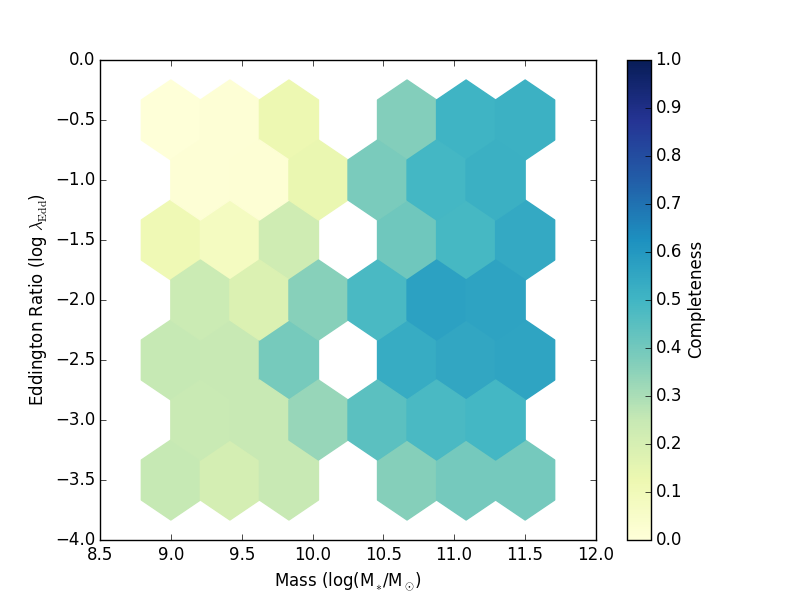}}
\end{subfigure}
\begin{subfigure}
	\centering
	\scalebox{0.39}
	{\includegraphics{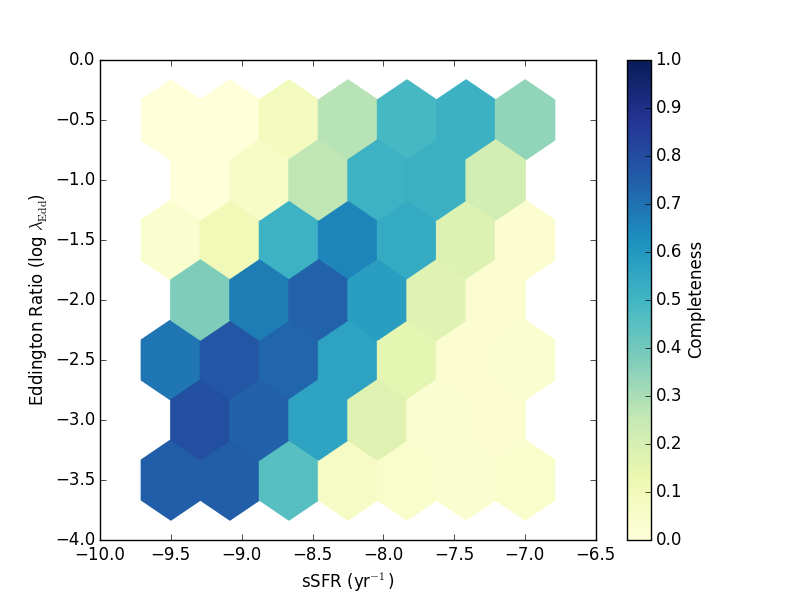}}
\end{subfigure}
\caption{Simulated parameters of the line ratio gradient technique, shaded by completeness, for a cut in the difference between the nuclear and extended [O III]/H$\beta$ ratio where $\Delta$log([O~III]/H$\beta$) $>$ 0.1. Compared to AGN selection using integrated line ratios (Figure~\ref{hex_diff_tot}), our spatially resolved line-ratio selection is able to find many AGN in low-mass and moderately star-forming galaxies.}
\label{hex_diff}
\end{figure}

\begin{figure}
\centering
\begin{subfigure}
	\centering
	\scalebox{0.39}
	{\includegraphics{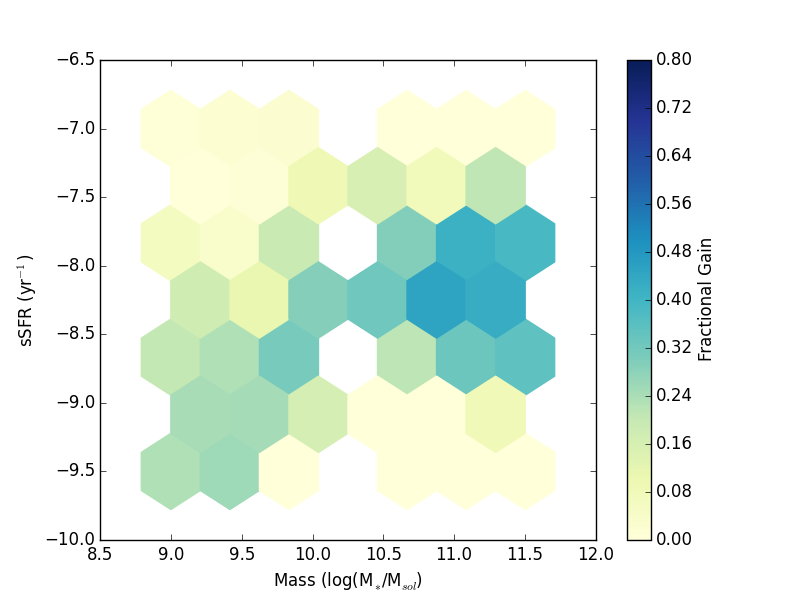}} \\
\end{subfigure}
\begin{subfigure}
	\centering
	\scalebox{0.39}
	{\includegraphics{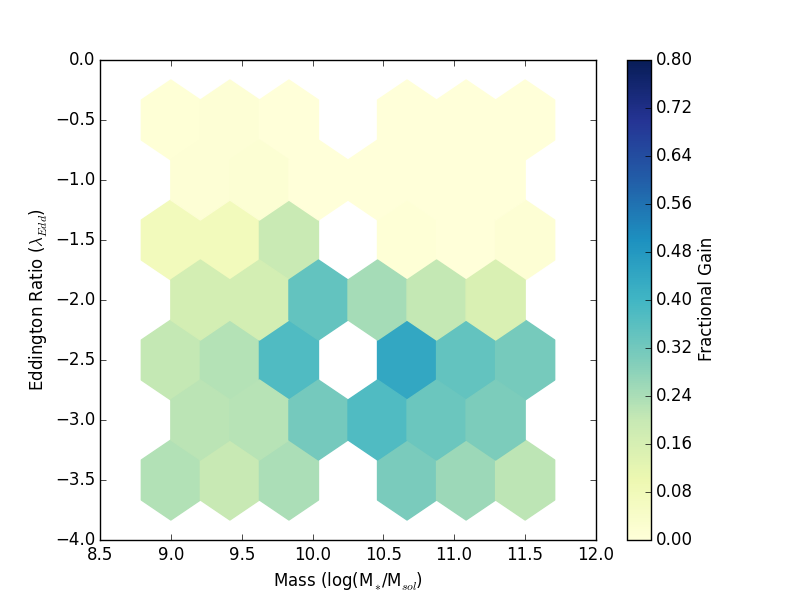}}
\end{subfigure}
\begin{subfigure}
	\centering
	\scalebox{0.39}
	{\includegraphics{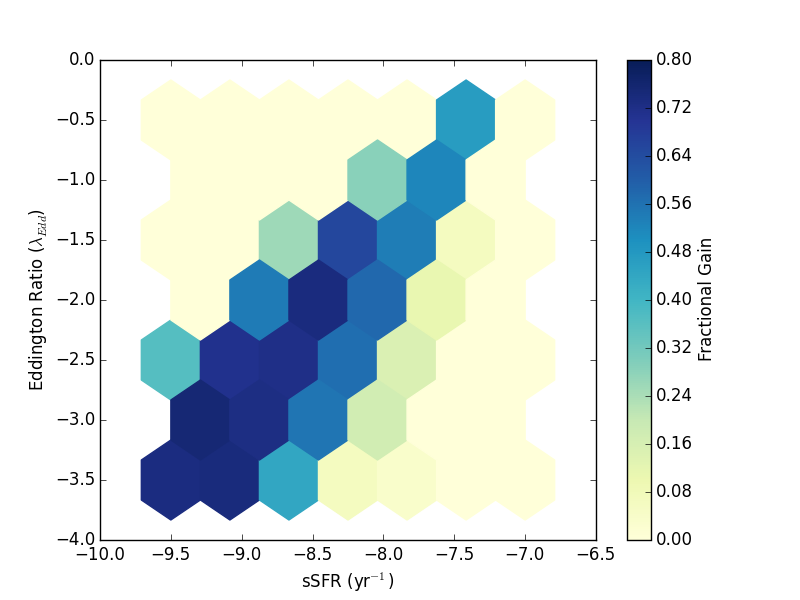}}
\end{subfigure}
\caption{The fractional gain in AGN identification when using the gradient technique.  The most gain occurs at lower mass when the AGN is low-luminosity and the star formation is low.  When used in conjunction with classical AGN identification methods, the gradient method finds many AGN that were previously unidentified.}
\label{gain}
\end{figure}

Figure~\ref{hex_diff} flattens one dimension of Figure~\ref{scatter3D}, showing the parameters in the simulations, shaded by completeness ($\Delta\log$([O~III]/H$\beta$) $> 0.1$) at each grid point. The gradient technique augments the parameter space when compared to Figure~\ref{hex_diff_tot}, particularly in the low sSFR, low Eddington ratio corner. This is demonstrated in Figure~\ref{gain}, which demontstrates the fractional gain in AGN identification. This technique is less effective at high sSFR and and low Eddington ratio, where the star formation emission swamps the signal from an under-luminous AGN.  Additionally, at very low masses, the S/N becomes low, but this effect can be mitigated by further stacking. However, using the gradient method identifies an average of $\sim 45\%$ more galaxies with AGN than via traditional line ratio methods, particularly at low Eddington ratio and low sSFR. 

\begin{figure}
\centering
\scalebox{0.8}
{\includegraphics{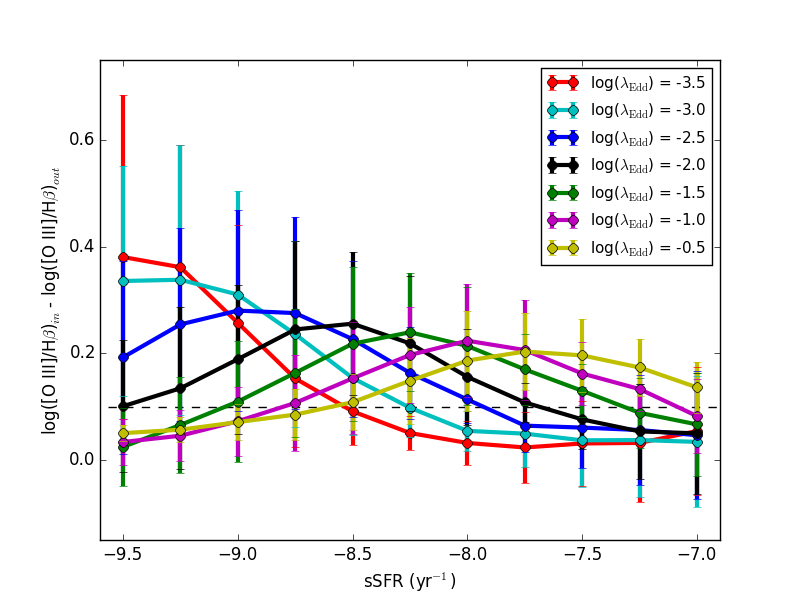}}
\centering
\caption{Gradient of log([O~III]/H$\beta)$ vs.\ sSFR for the highest stellar mass bin in the simulation.  The lines are shaded by Eddington ratio.  For the lowest sSFR, all AGN are detectable (where $\Delta$log([O~III]/H$\beta$) = 0.1 is the threshold shown with the dashed black line). As the sSFR increases, the lower Eddington ratio AGN are more difficult to detect, as the emission from disk star formation overwhelms the AGN\null. Note the subtle turnover at log(sSFR) $\sim -8$ yr$^{-1}$ where the PSF of the luminous AGN overwhelm weak extended star formation emission in low-sSFR galaxies, resulting in a flat gradient (and the integrated line ratio of an AGN). }
\label{ssfr_gradient}
\end{figure}

To illustrate the trends of the line ratio gradient with the various input parameters of the simulations, we have examined how the gradient changes with mass, sSFR, and Eddington ratio. Figure~\ref{ssfr_gradient} shows how the gradient shifts with sSFR, shaded by Eddington ratio in the highest mass bin (log(M$_*$) = 11.5). At the lowest sSFRs, the gradient method identifies the AGN for all Eddington ratios.  As the sSFR increases, the lowest Eddington ratio AGN simulations become overwhelmed by the star formation emission. 

Finally, we show in Figure~\ref{other_sims} how the simulation gradients change with redshift and morphology as compared to the original simulations.  To test different galaxy shapes, we performed simulations with $e=0.2$ as well as using a De Vaucouleur's $n=4$ profile instead of an exponential.  We also performed simulations at $z=1.4$ and $z=2.2$ to test the effects of different grism dispersion at the extreme wavelengths covered by G141 grism. We find that while there are some slight variations in gradient among the results, all differences are statistically insignificant (i.e., within the errors of each other.) 

\begin{figure}
\centering
\scalebox{0.8}
{\includegraphics{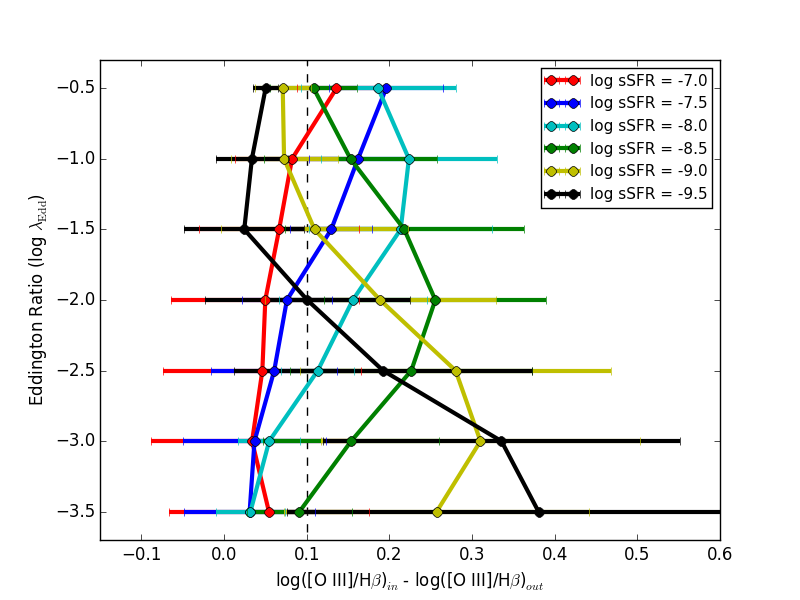}}
\centering
\caption{Eddington ratio vs.\ $\Delta$log([O~III]/H$\beta$) for the highest stellar mass bin in the simulations.  The lines are shaded by sSFR. The black dashed line is $\Delta$log([O~III]/H$\beta$) = 0.1. If the stellar mass and sSFR of a galaxy are known, this relationship suggests that it is possible to determine the Eddington ratio of the central AGN given the log([O~III]/H$\beta$) gradient.}
\label{gradient_edd}
\end{figure}

\begin{figure}
\centering
\begin{subfigure}
	\centering
	\scalebox{0.4}
	{\includegraphics{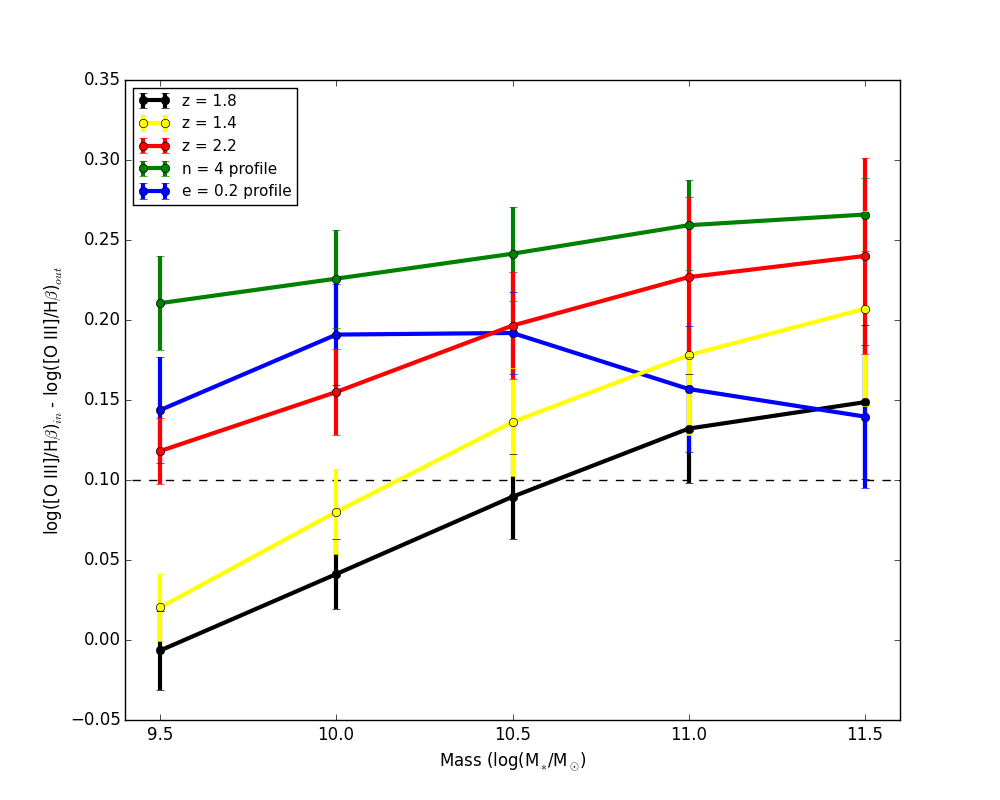}} \\
\end{subfigure}
\begin{subfigure}
	\centering
	\scalebox{0.4}
	{\includegraphics{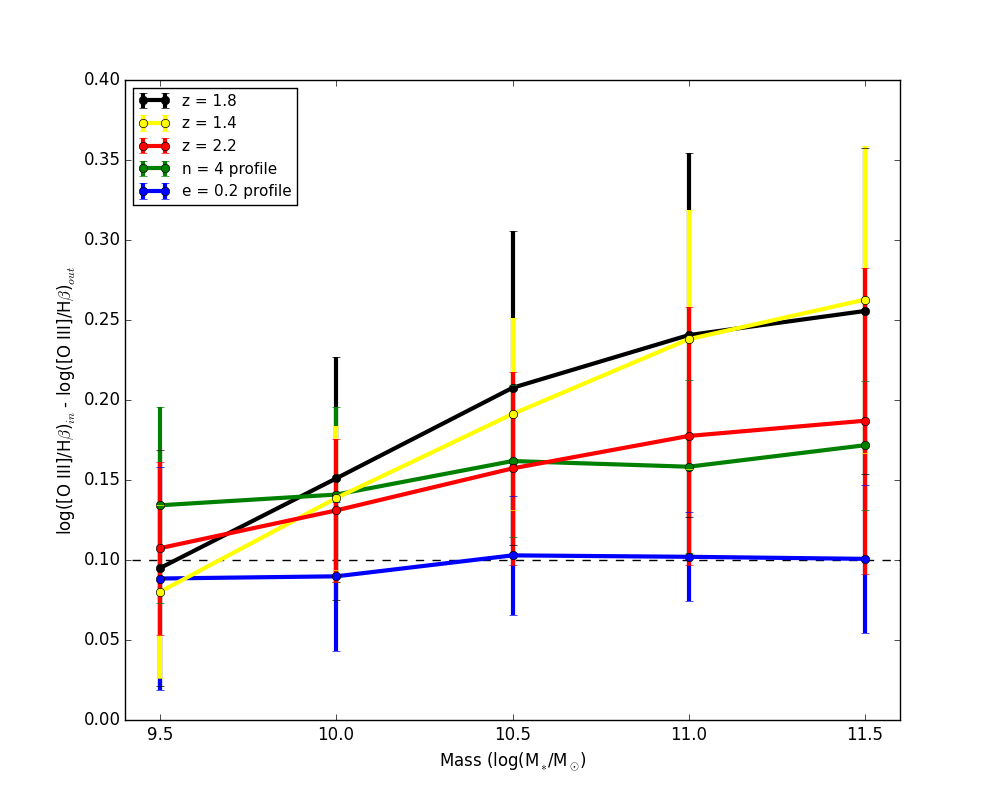}}
\end{subfigure}
\caption{\emph{Top:} Stellar mass vs. $\Delta$log([O~III]/H$\beta$) for various redshifts and morphology for log($\lambda_{\textrm{Edd}}) = -0.5$ over all sSFRs. The solid black line shows the $z=1.8$ simulations, while the red and yellow line shows the $z = 2.2$ and $z = 1.4$ simulation results, respectively.  The simulations of galaxies with $e=0.2$ are shown with the blue line and the green indicates the use of a DeVaucouleurs profile.  The dotted line indicates the $\Delta$log([O~III]/H$\beta$) = 0.1 cutoff. The simulations perform as well as or better than the full set at $z = 1.$ \emph{Bottom:} Same as top but with log(sSFR) = -.0 yr$^{-1}$, plotted over all Eddington ratios.}
\label{other_sims}
\end{figure}

The utility of the gradient method is emphasized when we compare the measured gradient to the Eddington ratio of the AGN\null.  Figure~\ref{gradient_edd} illustrates these trends for the highest stellar mass bin, with the lines shaded by sSFR. The conclusion is that for the and average sSFR, AGN of almost any Eddington luminosity are identified. That is the power of using line ratio gradients instead of total integrated line luminosities---low-luminosity or obscured AGN that were previously lost to traditional methods can now be found, enabling a much more complete picture of the AGN census during the height of star formation in the universe.

\begin{figure}
\centering
\scalebox{0.5}
{\includegraphics{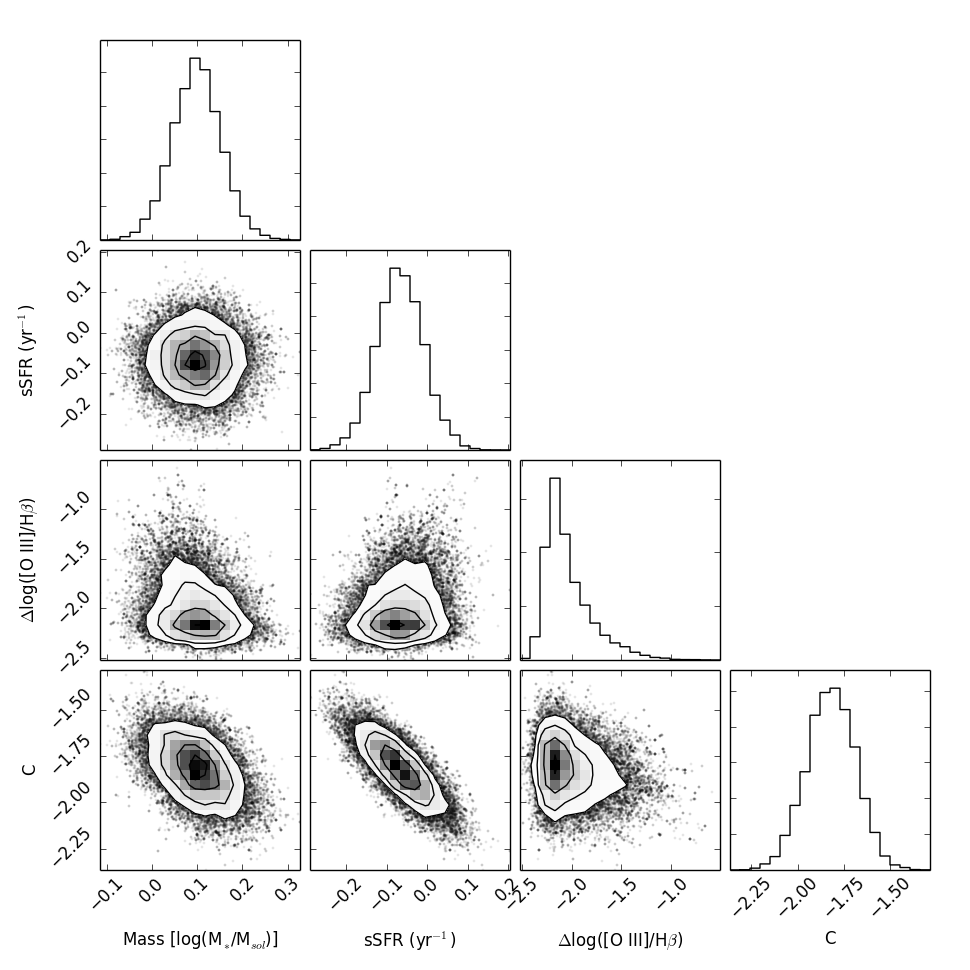}}
\centering
\caption{MCMC fitting results for the fit of a plane to mass, sSFR, and line ratio gradient to obtain an Eddington ratio for an AGN based on galaxy observables. The fit is well-constrained and the distributions for each parameter are relatively Gaussian. The parameter $C$ is a constant.}
\label{fit}
\end{figure}

We can apply the gradient technique to make predictions about the properties of the central AGN\null.  We fit a plane to the mass, sSFR, and ``observed" line-ratio gradient of the suite of simulated galaxies using the \texttt{emcee} Markov Chain Monte Carlo (MCMC) linear regression algorithm \citep{foreman2013}. This process produces a simple method for taking observed quantities and predicting the AGN Eddington ratio.  The fit is given by
\begin{equation}
\begin{split}
\log(\lambda_{\textrm{Edd}}) &  = (-1.82 \pm 0.13) - (0.099 \pm 0.053)[\textrm{log}(M_*) - 9]\\
& + (0.067 \pm 0.057)[\textrm{log(sSFR)}+10] - (2.10\pm0.21)\Delta \textrm{log([O~III]/H}\beta)\\
\end{split}
\end{equation}
The coefficients are all well-constrained, as shown in Figure~\ref{fit}. The intrinsic scatter of the relation is $\sigma = 0.30\pm0.22$.  The uncertainty in line-ratio gradient is a measurement error in the range  $\sim$ 0.1--0.3~dex, estimated in the simulations using \texttt{aXesim} for G141 spectra of two-orbit depth.  The errors in mass and sSFR are model uncertainties: In our simulations, these are the uncertainties in translating mass and sSFR to the ``observed" [O~III], H$\beta$, and continuum flux of the galaxy and AGN\null.  As model uncertainties, the mass and sSFR errors are not known, but are likely to be $\gtrsim$ 0.1~dex.  These combined uncertainties in the dependent variables dominated the scatter reported by the MCMC fit, so the intrinsic scatter of Equation 3 is not well-constrained.  For simplicity, we ignore the intrinsic scatter of Equation 3 when applying to the observations below, arguing that the uncertainties in mass, sSFR, and gradient dominate the resulting error in Eddington ratio (which is $\sim$ 1~dex).

We now have the ability to determine the Eddington ratio of an AGN for our simple galaxy model by measuring only one new quantity: the line ratio gradient of a galaxy. In the next section, we apply this to real data and show how our results are consistent with previous results. 

\section{Applications to Data}
We now apply our simulation results to real \emph{HST} WFC3 G141 grism observations.  In particular, we use the published line ratio gradients of \cite{trump2011, trump2014}, translating these observations into black hole accretion rate estimates.

\citet{trump2011} studied line ratio gradients in low-mass galaxies using G141 grism observations in the Hubble Ultra Deep Field (HUDF), taken as part of the supernovae follow-up in the CANDELS survey \citep{grogin2011, rodney2012}.  From the stacked spectrum of 28 galaxies with median M$_* = 10^{9.1}$M$_\odot$ and median $z=1.6$, \cite{trump2011} found a higher [O~III]/H$\beta$ ratio in the stacked spectrum core than in its extended region.  Coupled with the $L_X/L_{\textrm{[O III]}}$ ratio from stacked X-ray data, they concluded that ``at least some of these low-mass, low-metallicity galaxies harbor weak active galactic nuclei."  Armed with the results of our simulation, we can test this conclusion, quantifying the black hole growth in these dwarf galaxies.

We re-computed the line ratio gradients of the \citet{trump2011} stacked spectrum using the same collapsing method and inner/outer apertures used in our simulations, measuring a line ratio gradient of $\Delta$log([O III]/H$\beta$) = $0.312\pm0.129$. Given the median stellar mass (log(M$_*$/M$_\odot$) = 9.1) and specific star formation rate (log(sSFR [yr$^{-1}$]) = $-9.2$) of the sample and our derived Equation 3, the observed line ratio gradient implies an Eddington ratio of log($\lambda_{Edd}$) = $-2.43\pm0.31$. \cite{trump2011} posited that AGN were probably common in low-mass star-forming galaxies at $z\sim2$.  The calculated Eddington ratio for these stacked spectra indicates some presence of AGN, suggesting rapid black hole growth in these dwarf galaxies.  Deep X-ray stacking of 4~Ms Chandra observations similarly found evidence for widespread AGN with high Eddington ratios in low-mass dwarf galaxies \citep{xue2012}.

We additionally apply our simulation results to the observations of \citet{trump2014}, who measured line ratio gradients of stacked spectra of mass-matched samples of $z\sim2$ clumpy, smooth, and intermediate-morphology galaxies.  While the smooth and intermediate galaxies showed elevated [O~III]/H$\beta$ ratios in their centers, the clumpy galaxies did not, leading \citet{trump2014} to conclude that AGN are not preferentially fueled by clumpy galaxies.  Here we translate the measured line ratio gradients of the three galaxy samples into Eddington ratios.

As before, we begin by re-measuring the line ratio gradients of the \citet{trump2014} stacked spectra, using the annular-collapse method and the same inner and outer apertures as the simulations. Table~\ref{results} presents the observed line ratio gradients, median stellar masses, and specific star formation rates for each of the clumpy, intermediate, and smooth galaxy samples.  Also shown are the implied Eddington ratios calculated following our derived Equation 3. Our results indicate that the clumpy galaxies have slightly higher Eddington ratios, but are similar within the errors to the smooth  and intermediate galaxies from the sample, supporting the conclusion that AGN are not preferentially fueled by clumpy galaxies. Although individual clumpy galaxies have morphologies which differ from our simple simulations, the stacked spectra of clumpy galaxies are smooth and symmetric (see Figure 8 of \citet{trump2014}).  Thus our idealized simulations remain effective in interpreting the stacked observational data of clumpy galaxies.

\cite{mullaney2012} noted that there appears to be a constant ratio of black hole accretion to star formation rate, where $\dot{M}_{\rm BH}/{\rm SFR}=[0.5-0.7]\times 10^{-3}$ ($-3.30$ to $-3.15$ in log units) for $z\sim2$ galaxies.  From this, they draw the conclusion that there exists what they called an AGN ``main sequence." Using the Eddington ratios calculated above, we are able to examine if these AGN support the existence of this main sequence.  Arranging the relationship of black hole accretion and star formation rate into known quantities of sSFR and Eddington ratio, we obtain
\begin{equation}
\frac{\dot{M}_{\rm BH}}{\rm SFR} = \dfrac{\lambda_{\rm Edd}}{\rm sSFR} \dfrac{2.02 \times 10^{8}}{\eta  c^2}
\end{equation}
where we have assumed $\eta = 0.1$, $M_{\rm BH}/M_* = 1\times 10^{-3}$, sSFR in units of yr$^{-1}$, and $c$ in cgs units. The results are given in Table~\ref{results}.  All of the galaxies fall well below the average found by \citep{mullaney2012}. This is due to the lower Eddington ratios found for these stacks, indicating that they are not comprised of particularly powerful AGN.

\section{Discussion}
Comparison of Figures 7 and 2 demonstrates that our line-ratio gradient method is more complete and reliable than integrated line ratios.  In particular, the use of spatially determined line ratios allows for superior AGN identification in low-mass and high-sSFR host galaxies in which the emission lines associated with star formation overwhelm the AGN in integrated line ratios.  Our line-ratio gradient also goes beyond simple binary AGN classification, enabling basic estimates of AGN accretion rate (Equation 3).  Our simulations demonstrate that these line-ratio gradients have great utility for quantitative studies of black hole growth as a function of galaxy properties.

This work was specifically designed for the spatially distinguished [OIII]/H$\beta$ ratios in stacked spectra of $\sim20$ galaxies produced by the HST/WFC3 G141 grism, as in the 3D-HST survey \citep{brammer2012}.  However, our general method can be applied to any spatially resolved spectroscopy.  The key to using line-ratio gradients for AGN selection is contrast between the nuclear AGN emission lines and the extended emission associated with star formation.  Coarser spatial resolution, as in seeing-limited integral field unit (IFU) spectroscopy surveys of $z>1$ galaxies from the ground \citep[e.g.,][]{forster2009, wisnioski2015}, will offer lower efficiency for AGN selection with line ratio gradients.  Similarly, the lower spatial resolution ($~0$\farcs2) and lower sensitivity of the Euclid grisms will result in slightly lower completeness for AGN identification, although the Euclid surveys will include a much larger sky area than 3D-HST \citep{laureijs2014}.

Meanwhile increasing the contrast between nuclear AGN and extended star formation will provide increased sensitivity for AGN selection.  For example, using [O~III]/H$\beta$ with the \emph{HST}/WFC3 G102 grism is likely to be more complete to AGN selection: The G102 grism has the same spatial resolution as G141 but accesses [O~III]/H$\beta$ in lower-redshift galaxies that are consequently larger and more extended compared to a point-source AGN\null.  The slitless grism of the Near-InfraRed Imager and Slitless Spectrograph (NIRISS) on the \emph{James Webb Space Telescope} (\emph{JWST}) will have slightly lower spatial resolution (0\farcs2 $\times$ 0\farcs46), suggesting lower AGN completeness using a single line ratio.  However \emph{JWST} also offers higher spectral resolution to reach multiple line ratios (e.g. [N~II]$\lambda6585$/H$\alpha$, [Ne~III]$\lambda369$/[O~II]$\lambda3727$) and dramatically increased sensitivity for access to weak, high-ionization lines like [Ne~V] $\lambda3426$ and He~II$\lambda\lambda1640,466$.

We chose to focus on the [O~III]/H$\beta$ line ratio because these are some of the strongest lines available to the \emph{HST}/WFC3 grism.  The [N~II]/H$\alpha$ and [S~II]$\lambda\lambda671,6732$/H$\alpha$ ratios are similarly strong and offer a slightly different perspective with their lower-ionization forbidden lines.  However, the [N~II] and H$\alpha$ are indistinguishably blended in the low-resolution \emph{HST} grisms. Spatially resolved spectroscopy with higher-resolution access to [NII]/H$\alpha$ and [SII]/H$\alpha$ offer a different perspective on AGN selection: this includes current ground-based IFU spectroscopy \citep[e.g.,][]{genzel2014} and future \emph{JWST} spectroscopy.  The [Ne~III]/[O~II] line ratio is another intriguing alternative: Although both lines are weaker, they are detected in stacked G141 grism spectra that indicate that $z\sim2$ galaxies have a mysterious over-abundance of [Ne~III] \citep{zeimann2015}.  High-redshift galaxies also have strangely high C~IV$\lambda$1550 and C~III]$\lambda$1909 equivalent widths \citep{stark2014, stark2015}, and with sufficient sensitivity (e.g., from \emph{JWST}/NIRSpec), the spatially resolved C~IV/C~III] ratio might provide a unique handle on AGN selection in the highest-redshift ($z>5$) galaxies.

\section{Summary}
We have presented a new grism-based technique for identifying low-luminosity AGN in stacked HST/WFC3 grism spectra of $z\sim2$ galaxies by determining the line ratio gradient between the nuclear and extended regions.  To explore the utility of this method, we simulated galaxies with various disk star formation and nuclear activity and measured how the ratio gradient is affected by varying these parameters. Our method allows us to detect lower luminosity or obscured AGN that are not detectable by traditional methods that use total integrated line luminosities (i.e., the BPT and MEx diagrams). In applying our new method to real data, we have confirmed previous conclusions drawn about AGN and their hosts by \cite{trump2011, trump2014} and \cite{mullaney2012}. The consistency of our Eddington ratio estimates with previous stacking of observed spectra validates the simple galaxy model used in our simulations.

For this paper we have used exponential disks to model the extended region of the galaxy: this is generally appropriate for stacked spectra of star-forming galaxies.  In the future, these simulations can be modified to include parameters such as morphology using, for example, the S\`ersic index. Additionally, while we have separated the galaxy into the nuclear and extended regions, a finer scale could be used to capture further nuances in how emission changes across the galaxy. This gradient technique can also be expanded to encompass a wide range of science applications, such as calculating the metallicity gradient across a galaxy. 

We thank the anonymous referee for their very insightful comments. This work was partially supported by NSF through grant AST 09-26641. JRT acknowledges support from NASA through Hubble Fellowship grant \#51330 awarded by the Space Telescope Science Institute, which is operated by the Association of Universities for Research in Astronomy, Inc., for NASA under contract NAS 5-26555. The Institute for Gravitation and the Cosmos is supported by the Eberly College of Science and the Office of the Senior Vice President for Research at The Pennsylvania State University. IRAF is distributed by the National Optical Astronomy Observatory, which is operated by the Association of Universities for Research in Astronomy (AURA) under a cooperative agreement with the National Science Foundation. This research has made use of NASA's Astrophysics Data System Bibliographic Services.

\newpage
\bibliography{grism_sim}

\newpage

\begin{deluxetable}{ccc}
\tablecolumns{3}
\tablewidth{0pc}
\tablecaption{Parameter Space of Simulated Galaxy Properties \label{tbl-1}}
\tablehead{ \colhead{(log($M_*$ [M$_\sun$])} & \colhead{log(sSFR [yr$^{-1}$])} & \colhead{log($\lambda_{\textrm{Edd}}$)}}
\startdata
9.0 & $-9.50$ & $-3.5$ \\
9.5 & $-9.25$ & $-3.0$ \\
10.0 & $-9.00$ & $-2.5$ \\
 10.5 & $-8.75$ & $-2.0$\\
11.0 & $-8.50$ & $-1.5$ \\
11.5 & $-8.25$& $-1.0$\\
 - & $-8.00$ & $-0.5$\\
 - &  $-7.75$ & - \\
 - &  $-7.50$ & -  \\
  - & $-7.25$ & - \\
  - & $-7.00$& - \\
\enddata
\end{deluxetable}

\begin{deluxetable}{cccccc}
\tabletypesize{\footnotesize}
\tablecolumns{6}
\tablewidth{0pc}
\tablecaption{Eddington Ratios\label{results}}
\tablehead{&\colhead{(log($M_*$ [M$_\sun$])} & \colhead{log(sSFR [yr$^{-1}$])} & \colhead{$\Delta$log([O III]/H$\beta$)}&\colhead{$\lambda_{\textrm{Edd}}$ }&\colhead{log($\dot{M}_{\rm BH}/{\rm SFR}$})}
\startdata
HUDF & $9.1$ & $-9.2$ & $0.312\pm0.129$ & $-2.43\pm0.31$ & $-4.88\pm0.33$ \\
Clumpy & $9.2$ & $-8.49$ & $-0.085\pm0.140$\tablenotemark{a} & $-1.80\pm0.34$ & $-4.96\pm0.35$ \\
Intermediate & $9.71$ &  $-8.45$ & $0.128\pm0.094$ & $-2.06\pm0.27$ & $-5.26\pm0.29$ \\
Smooth & $9.78$ & $-8.51$ &  $0.185\pm0.230$ & $-2.19\pm0.52$ & $-5.33\pm0.53$\\
\enddata
\tablenotetext{a}{A negative gradient is treated as 0.0}
\end{deluxetable}

\end{document}